# Pressure Study of Superconducting Oxypnictide LaFePO

Kazumi IGAWA, Hironari OKADA, Kazunobu ARII, Hiroki TAKAHASHI, Yoichi KAMIHARA[1], Masahiro HIRANO[1,2], Hideo HOSONO[1,2], Satoshi NAKANO[3] and Takumi KIKEGAWA[4]

*Department of Physics, College of Humanities & Sciences, Nihon University, Sakurajousui, Setagaya-ku, Tokyo 156-8550*

*[1]ERATO-SORT, JST, in Frontier Research Center, Tokyo Institute of Technology, 4259 Nagatsuda, Midori-ku, Yokohama 226-8503*

*[2]Frontier Research Center, Tokyo Institute of Technology, 4259 Nagatsuda, Midori-ku, Yokohama 226-8503*

*[3]National Institute for Material Science, Namiki, Tsukuba, Ibaraki 305-0044, Japan*

*[4]Institute of Material Structure Science, High Energy Accelerator Research Organization, Oho, Tsukuba, Ibaraki 305-0801, Japan*

Electrical resistivity and magnetic susceptibility measurements under high pressure were performed on an iron-based superconductor LaFePO. A steep increase in superconducting transition temperature ($T_c$) of LaFePO with $dT_c/dP > 4$ K/GPa to a maximum of 8.8 K for $P = 0.8$ GPa was observed. These results are similar to isocrystalline $LaFeAsO_{1-x}F_x$ system reported previously. X-ray diffraction measurements were also performed under high pressure up to 10 GPa, where linear compressibility $\kappa_a$ and $\kappa_c$ are presented.

The Fe-based LaFePO[1,2] and the Ni-based LaNiOP[3] have recently been reported to undergo superconducting transition with $T_c$ of 4 and 3 K, respectively. Furthermore, the isocrystalline F-doped LaFeAsO was discovered to show the $T_c$ of 26 K[4], which had a significant impact in the field of condensed matter physics. Right after the discovery of the superconductivity of F-doped LaFeAsO, applying pressure on it increases $T_c$ to 43 K[5], which is the highest $T_c$ reported so far, with the exception of cuprate high-$T_c$ superconductors. These results suggest that the compression is advantageous means to investigate superconductivity. Indeed, the iron-pnictides with a $T_c$ above 50 K was reported by substituting smaller rare earth elements, such as Nd, Sm etc., for La atom.[6,7] This indicates the chemical pressure is effective to increase $T_c$.

The crystal structure of LaFePO comprises a stack of alternation LaO and FeP layers. It is thought that LaO layer is insulating whereas FeP layer is conductive nature. The conductive carriers are confined two dimensionally in the FeP layers[8], whose concentration can be controlled by oxygen content or atomic substitution in the LaO layer. This situation is similar to that in cuprate high-$T_c$ superconductors, where an enhancement in the $T_c$ has been observed with increasing carriers in a conductive $CuO_2$ layer through chemical doping into the block layer. Moreover, various types of pressure effects on $T_c$ have been reported in the cuprate high-$T_c$ superconductors[9], since the pressure often increases the carrier concentration in the $CuO_2$ plane through changing the charge distribution due to anisotropic compression. However, in $LaFeAsO_{1-x}F_x$, little is known about the mechanism of such a large enhancement of $T_c$ by pressure because of the lack of structural data, compared with cuprate high-$T_c$ superconductors. In LaFePO, various doping effects indicates the shrinkage of lattice constant $c$ increases $T_c$, while there is no correlation of $T_c$ with lattice constant $a$.[10] The lattice constant $c$ ranging from 0.8520 to 0.8500 nm corresponds to the $T_c$ from 2 to 7 K in LaFePO at atmospheric pressure. The pressure effect on $T_c$ for LaFePO above 5 GPa has been reported[11] so far. However, there reported no data below 5 GPa. Since the quite large pressure effect on $T_c$ was observed below 5 GPa in $LaFeAsO_{1-x}F_x$, it should be important for LaFePO to reveal the high pressure effect on $T_c$ more precisely in the low pressure range.

In this study, electrical resistivity and magnetic susceptibility measurements under high pressure up to 12 GPa are performed to obtain the pressure dependence of $T_c$, and x-ray diffraction up to 10 GPa is performed to examine the relation between superconductivity and crystal structure. Another advantage of high-pressure experiment is free from a side effect, such as impurities or oxygen defects often observed in the substitution experiments.

Polycrystalline LaFePO was prepared by the solid-state reactions reported previously.[1,10]

Electrical resistivity measurements under high pressure were performed by means of a standard dc four-probe method. Pressures up to 1.5 GPa were applied and clamped at room temperature using a WC piston and NiCrAl cylinder device. A liquid pressure-transmitting medium (Fluorinert FC70:FC77=1:1) was used to maintain hydrostatic conditions. A diamond anvil cell (DAC) made of CuBe alloy was used for electrical resistivity measurements at pressures up to 30 GPa. In the case of DAC, the sample chamber comprising a rhenium gasket was filled with powdered NaCl as the pressure-transmitting medium, and thin (5-μm thick) platinum ribbons were inserted into the sample chamber to act as leads for the standard dc four-probe analysis. A thin BN layer acted as an electric insulation between the leads and the rhenium gasket. Fine ruby powder scattered in the sample chamber was used to determine the pressure by a standard ruby fluorescence method. The high-pressure magnetization measurements were performed by means of a SQUID magnetometer (Quantum Design MPMS) used in conjunction with a DAC made of CuBe and a liquid pressure-transmitting medium (methanol:ethanol=4:1). The high-pressure x-ray diffraction measurements were performed using synchrotron radiation at KEK-BL18C with the wavelength of 0.061642 nm. The DAC and a liquid pressure-transmitting medium (methanol:ethanol=4:1) were used for x-ray diffraction measurements.

Figure 1(a) shows the resistivity of LaFePO as a function of the temperature for $P$ < 1.5 GPa, using the piston cylinder device. The $T_c$ determined at the onset temperatures designated by an arrow in Fig. 1, where the $T_c$ increases rapidly from 5.8 K to 8.8 K with pressure at a rate of 4.0 K/GPa up to 0.8 GPa. However, the $T_c$ determined at zero resistivity increased more rapidly than the onset $T_c$, resulting in a sharpening of the superconducting transition with increasing pressure. This is unlikely to be due to a distribution of $P$ in the pressure cell, and is more probably caused by sample imhomogeneity, because the hydrostatic conditions were kept unchanged during the measurements. The onset $T_c$ decreased slowly above 0.8 GPa, and with a further increase in $P$ up to 12 GPa using DAC, it continues to decrease slowly. The superconducting transition becomes broader with increasing pressure in the DAC measurements (Fig. 1(b)), which is caused by a distribution in the nonhydrostatic compressive stress as a result of the use of solid pressure-transmitting medium. In another measurement (run 3 in fig.3) using DAC, although a zero resistance was not observed because of the nonhydrostatic condition, the onset $T_c$ shows the almost same value as other measurements. The decrease in the $T_c$ with $P$ was also confirmed by the curves of the magnetization normalized at 8 K ($M/M_{8K}$) against $T$ at $P$ < 7 GPa (Fig. 2). The onset $T_c$ in the magnetization-$T$ curve decreases, at almost the same rate as that of the $T_c$ in the resistance-$T$ curve. The diamagnetic signal accompanied with the superconducting transition was not observed above 7 GPa, which may be due to the rapid decrease of a volume fraction of superconductivity. Indeed, the resistance loss accompanied with superconducting transition tends to be small gradually at $P$ > 5 GPa. Fig. 3 summarizes the $T_c$ values in the $P$-$T$ phase diagram of LaFePO. Note that the initial increase rate (4.0 K/GPa) is larger than those of the cuprate high-$T_c$ superconductors, where the d$T_c$ /d$P$ values are about 1～2 K/GPa. In case of $LaFeAsO_{0.89}F_{0.11}$[5)], it shows much steeper increase in $T_c$ at a rate of 8 K/GPa. For both iron pnictides, the initial increase of $T_c$ with applying pressure is steeper than any other cuprate high-$T_c$ superconductors. On the other hand, J.J.Hamlin et al.[11)] reported the $P$-$T$ phase diagram for LaFePO above 5 GPa by measuring the electrical resistance using the DAC, where the $T_c$ decreases above 5 GPa. Though the $T_c$ of their sample shows a little higher value, it is almost consistent with our results.

Figure 4(a) shows the normalized lattice constants $a/a_0$ and $c/c_0$ of LaFePO at room temperature for $P$ < 10 GPa, using the DAC. The lattice along $c$ axis is more compressible than along $a$ axis. The linear compressibility $\kappa_a$ and $\kappa_c$ are $2.74 \times 10^{-3}$ and $4.95 \times 10^{-3}$ $GPa^{-1}$, respectively, which are calculated using the Murnaghan equation of state. The compressibility $\kappa$ was calculated to be $10.4 \times 10^{-3}$ $GPa^{-1}$. Fig. 4(b) shows the value of $c/a$ decreases linearly with applying pressure, which is usually observed in the layered compounds, for example, $\kappa_c$ is almost twice as $\kappa_a$ in $YBa_2Cu_3O_{7-y}$[12)] and $HgBa_2CaCu_2O_{6+\delta}$.[13)] These values are similar to the present results. On the other hand, the $T_c$ is reported to change from 2 to 7 K when atomic substitution is performed and/or oxygen defect is introduced to LaFePO.[10)] In these experiments, the $T_c$ increases with decreasing the lattice constant $c$, while the lattice constant $a$ does not correlate $T_c$ well. The lattice constant $c$ ranging from 0.8500 to 0.8520 nm corresponds to the change of $T_c$ from 2 to 7 K. Under high pressure, the change of the lattice constant $c$ from 0.8520 to 0.8500 nm corresponds to the application of pressure about 0.7 GPa, which raises the $T_c$ about 3 K.

Concerning the pressure effect on $T_c$, the iron pnictides superconductors, LaFePO and $LaFeAsO_{1-x}F_x$, have a common feature with the cuprate superconductors. Both compounds show the optimum $T_c$ under high pressure. In the cuprate superconductors, since the external pressure is considered to change the doping level through the anisotropic compression, the electronic state having the maximum $T_c$ under high pressure is regarded as the optimum doped. On the other hand, since the doping with $F^-$ ions enhances $T_c$ in the undoped LaFeAsO, it seems reasonable to suppose that the large enhancement of $T_c$ with applying pressure is caused by the change of the doping level of the electronic state of FeP layer through the anisotropic compression. In LaFePO, although a bulk modulus is 1.5 times larger than that of $LaFeAsO_{1-x}F_x$[14], the pressure dependence of the linear compressibility ratio $\kappa_c/\kappa_a$ shows the almost same value for both compounds. Then it is suggested that the pressure dependence of $T_c$ for LaFePO might be caused by the the change of the doping level of the electronic state in the FeP layer.

The superconducting phase diagram up to 12 GPa has been established for LaFePO by the electrical resistivity and magnetic susceptibility measurements. The $T_c$ increases steeply, and decreases through the peak value of 8.8 K at 0.8 GPa. This behavior is similar to the pressure dependence of $T_c$ in $LaFeAsO_{1-x}F_x$. On the other hand, it is likely that the change of the electronic state of the FeP layer is caused through the anisotropic compression obtained by the x-ray diffraction measurements. In order to study more precisely the electronic state under high pressure, it is necessary to determine the precise crystal parameters , such as each bond length and bond angle etc. as soon as possible. The high-pressure studies for F-doped LaFePO are now in progress to compare the pressure effect with the doping effect.

**Figure captions**

Fig.1. (Color online) (a) Temperature dependence of the electrical resistivity of LaFePO below 1.5 GPa, using piston cylinder device. The onset $T_c$ value is indicated by an arrow. (b) Temperature dependence of the electrical resistance of LaFePO at 3 and 5 GPa, using DAC.

Fig.2. (Color online) Temperature dependence of the magnetization normalized at 8 K of LaFePO up to 7 GPa using DAC and SQUID magnetometer.

Fig.3. (Color online) Superconducting phase diagram of LaFePO. The $T_c$ is defined at the onset of the transition. Run 1(open triangle) is the electrical resistivity measurements using piston cylinder device, Run 2(close triangle) and run 3(open circle) are runs using DAC. Run 4(close circle) in the inset is obtained from the magnetic susceptibility measurements.

Fig. 4. (Color online) (a) Pressure dependence of the lattice constants *a*(open circle) and *c*(close circle). The lattice along *c* axis is more compressible than along *a* axis. (b) Pressure dependence of the value of *c*/*a*. The *c*/*a* decreases linearly with applying pressure.

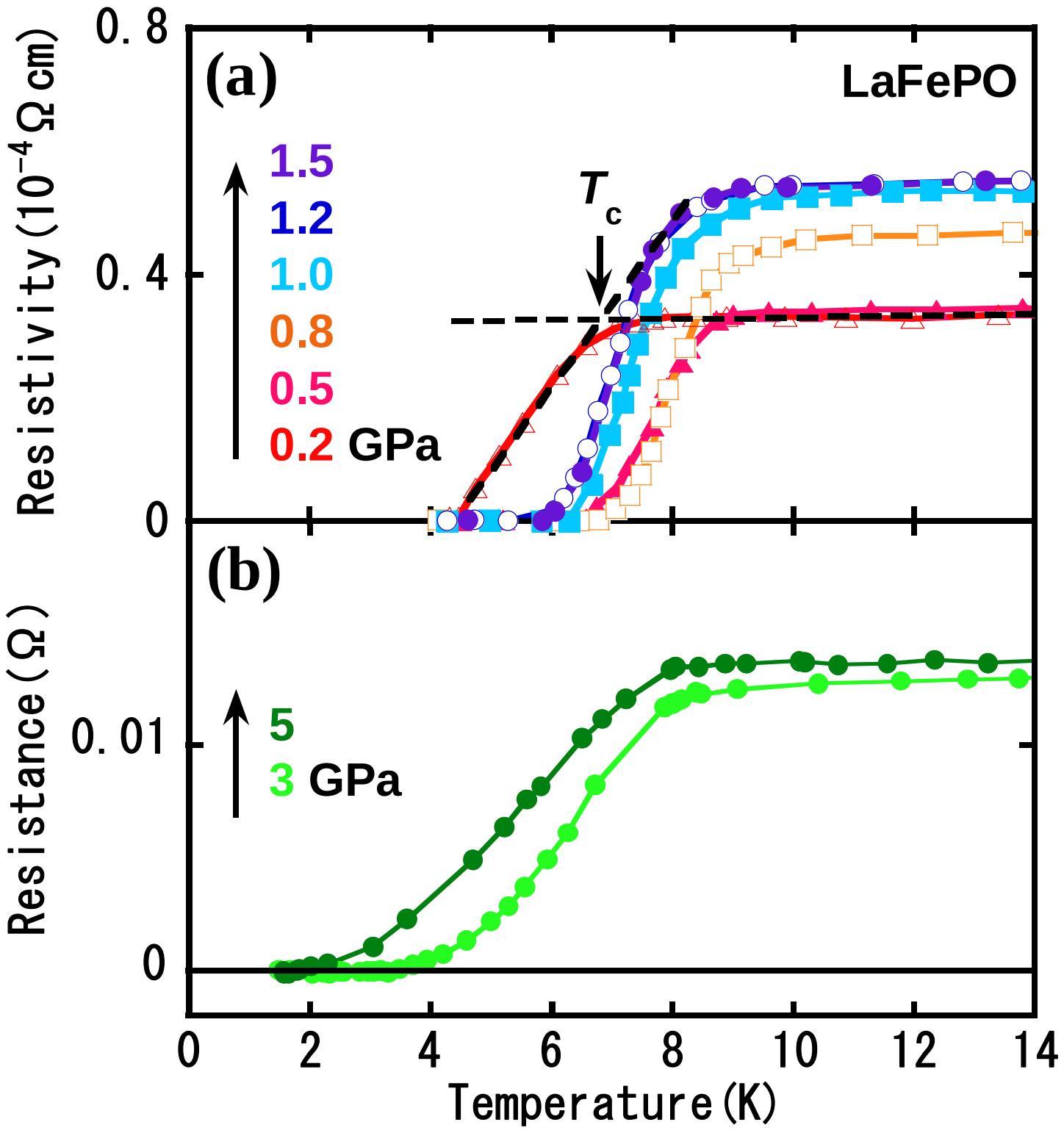

Fig. 1

H.TAKAHASHI et al.

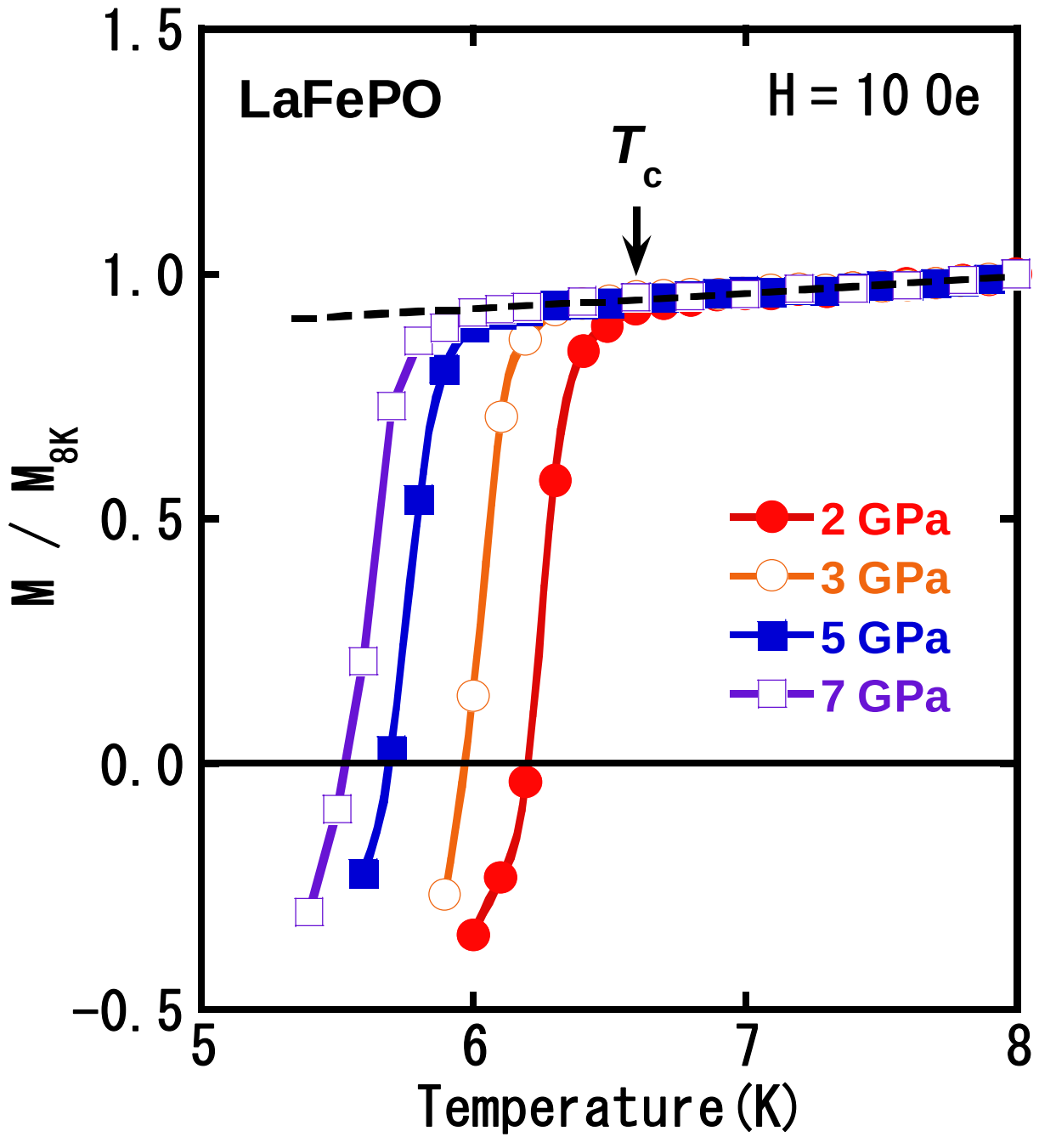

Fig. 2

H.TAKAHASHI et al.

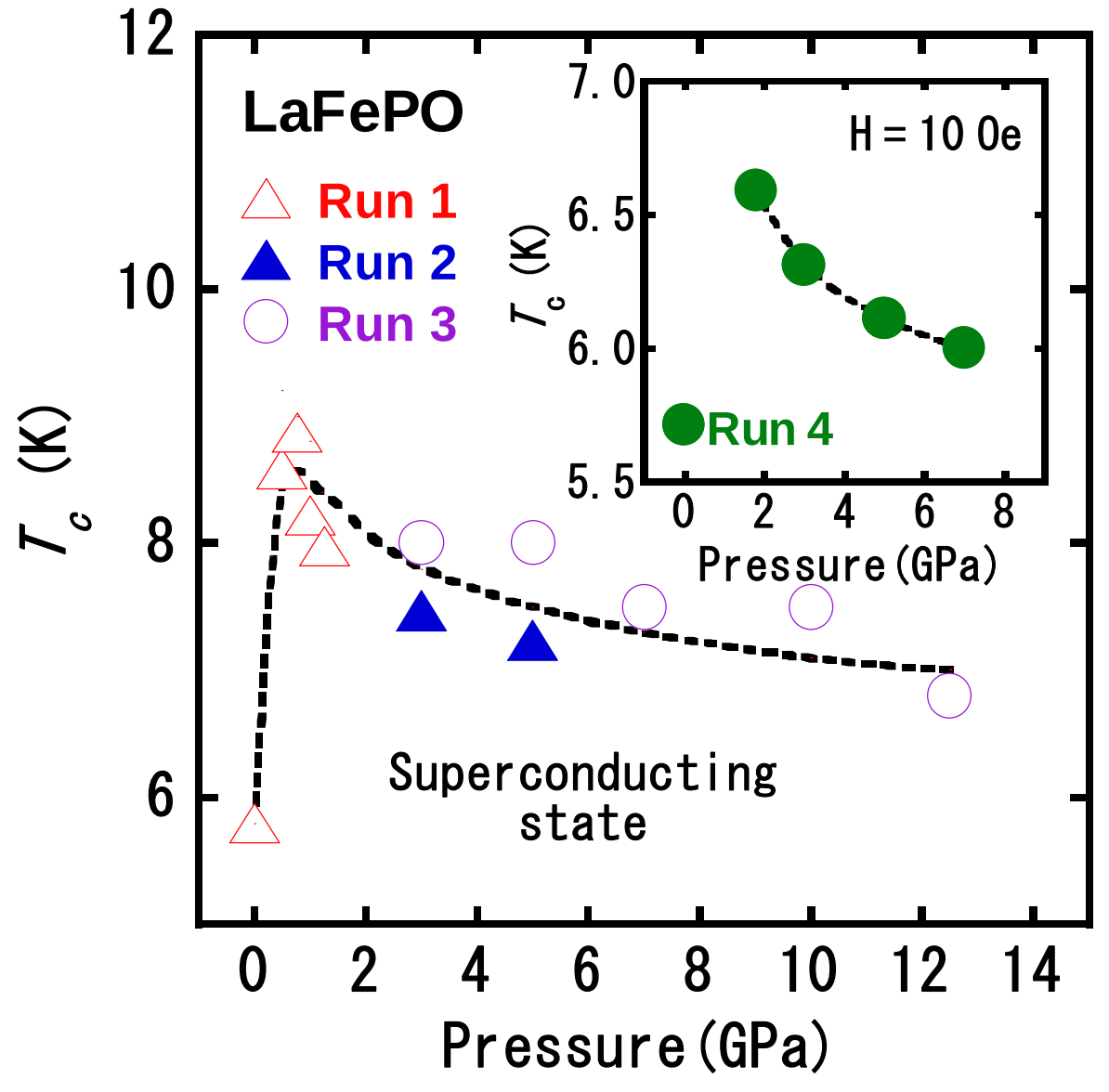

Fig. 3

H.TAKAHASHI et al.

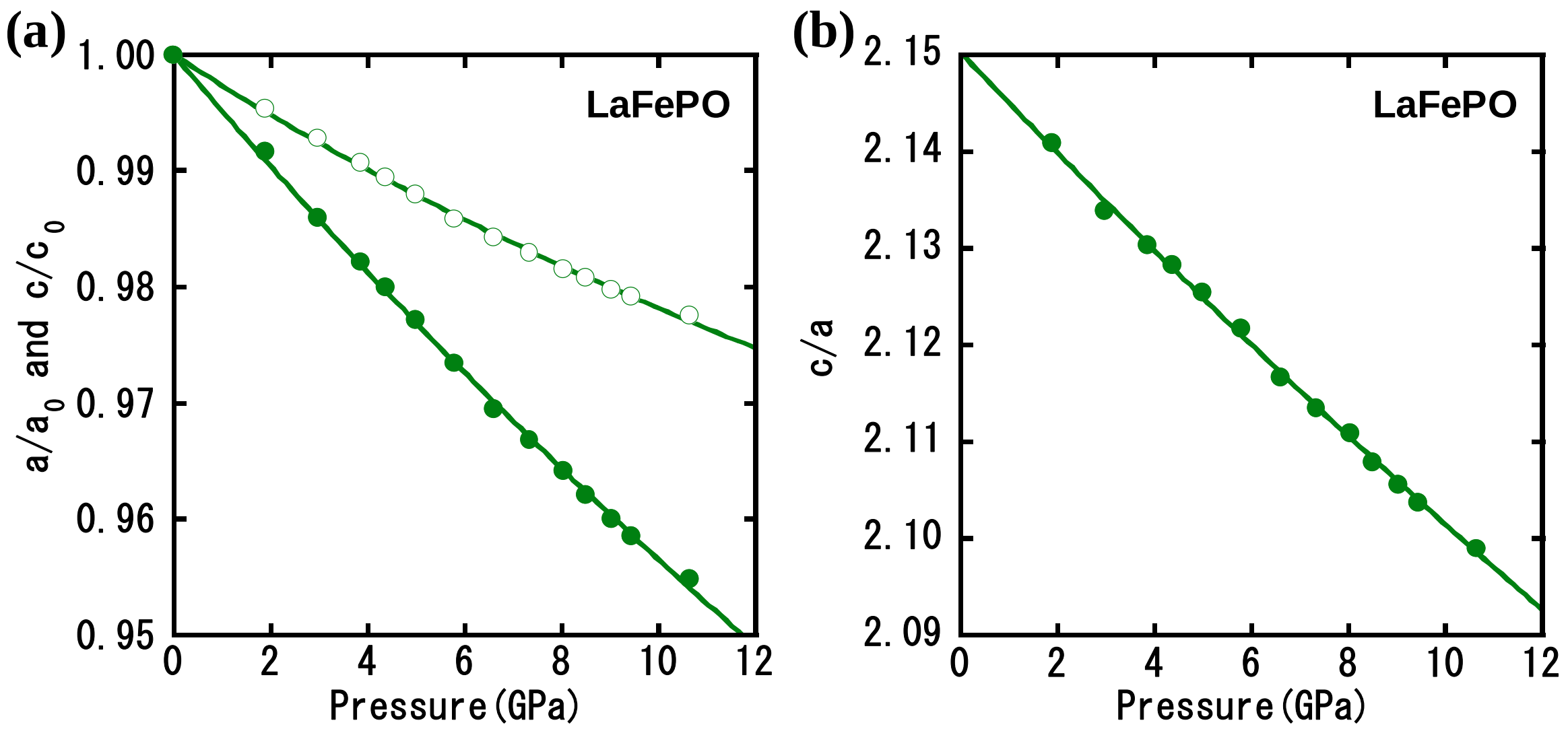

Fig. 4

H.TAKAHASHI et al.